# Origin of the tiny energy gap and Dirac points in monoclinic trilayer nickelate La$_4$Ni$_3$O$_{10}$

Hu Zhang

*Hebei Research Center of the Basic Discipline for Computational Physics, Hebei Key Laboratory of High-precision Computation and Application of Quantum Field Theory, College of Physics Science and Technology, Hebei University, Baoding 071002, People's Republic of China*

E-mails: zhanghu@hbu.edu.cn

Superconductivity was recently found in trilayer nickelate La$_4$Ni$_3$O$_{10}$ under high pressure with a phase transition from the monoclinic *P*2$_1$/*a* structure to the tetragonal *I*4/*mmm* structure. Previous experimental works have confirmed the existence of a tiny energy gap formed with Ni 3$d_{z^2}$ orbitals in monoclinic La$_4$Ni$_3$O$_{10}$. Here we investigate the physical origin of this gap by analyzing symmetry properties of energy bands based on the group theory. The tiny gap comes from energy bands with opposite parity at the Brillouin zone center. In addition, we also find previously unknown Dirac points in some momentum directions around the Fermi level. An effective Hamiltonian is constructed to describe low energy physics of the tiny energy gap and Dirac points. Due to the low crystal symmetry of monoclinic La$_4$Ni$_3$O$_{10}$, its energy bands display strong anisotropic properties.

## I. INTRODUCTION

Since the discovery of superconductivity in $Nd_{0.8}Sr_{0.2}NiO_2$ thin film in 2019 [1], layered nickelate oxides have attracted more and more attention. Recently, superconductivity near 80 K was found in Ruddlesden-Popper nickelate $La_3Ni_2O_7$ under high pressure [2]. There existed a structure transition from the orthorhombic (*Amam*) structure to the orthorhombic (*I4/mmm*) structure when $La_3Ni_2O_7$ is compressed [3]. Similar to $La_3Ni_2O_7$, superconductivity was also found very recently in trilayer nickelate $La_4Ni_3O_{10}$ under high pressure [4,5]. A phase transition from the monoclinic *P*2$_1$/*a* (or *P*2$_1$/*c* with a different choice of axis) to the tetragonal *I*4/*mmm* structure was confirmed. The electronic structure difference between different crystal structures in phase transitions is important to understand the mechanism of superconductivity in nickelates. For both $La_3Ni_2O_7$ and $La_4Ni_3O_{10}$, the Ni $3d_{z^2}$ orbital is believed to play a critical role for the appearance of superconductivity [2].

For $La_4Ni_3O_{10}$ with the monoclinic *P*2$_1$/*a* space group, a 20 meV energy gap near the Fermi level was found by angle-resolved photoemission spectroscopy (ARPES) measurements and theoretical band structure calculations [6,7]. This energy gap is formed by the band of principally Ni $3d_{z^2}$ orbital, which is very important to understand physical properties of monoclinic $La_4Ni_3O_{10}$. However, the origin of such energy gap is not well investigated up to now. To fully understand electronic structures of $La_4Ni_3O_{10}$, we should find out the physical origin of this tiny energy gap.

In this work, we firstly analyze the symmetry of electronic energy bands of monoclinic $La_4Ni_3O_{10}$ with the *P*2$_1$/*a* space group detailly with the help of group theory. Then, the origin of the energy gap is discussed. A low energy effective model is constructed to describe the low energy physics of energy gap based on the theory of invariant. In addition, through careful analysis of energy bands in the entire Brillouin zone, we also discover previously unrecognized Dirac points formed by Ni $3d_{z^2}$ orbital contributed bands near the Fermi level in some momentum directions.

## II. METHODOLOGY

We have performed first-principles calculations based on density functional theory

(DFT) [8] with the Perdew–Burke–Ernzerhoff (PBE) functional in generalized gradient approximation (GGA) [9] in the Vienna Ab Initio Simulation Package (VASP) [10-12]. We used a 12×12×12 Monkhorst-Pack grid [13] and an energy cutoff of 500 eV. The $k \cdot p$ method was used to obtain the effective Hamiltonian.

## III. RESULTS AND DISCUSSION
### A. Symmetry properties of monoclinic $La_4Ni_3O_{10}$

The space group number of monoclinic $La_4Ni_3O_{10}$ is 14. Due to the different choice of axis, the space group is denoted as $P2_1/a$ or $P2_1/c$. Recent experiments [4,14] for monoclinic $La_4Ni_3O_{10}$ with space group $P2_1/a$ at ambient pressure give $a$ = 5.4164 Å, $b$ = 5.4675 Å, $c$ = 14.2279 Å, α = β = 90°, and γ = 100.752°. For the notation of $P2_1/c$, $a$ = 14.2279 Å, $b$ = 5.4675 Å, $c$ = 5.4164 Å, α = γ = 90°, and β = 100.752°. To simplify the analysis of energy bands, we use the setting of $P2_1/c$ space group in this work. Different choices do not change the physics. In Fig. 1(a) we show atomic structures of primitive cell of monoclinic $La_4Ni_3O_{10}$ with the $P2_1/c$ space group. There are two formulas containing six Ni atoms in the primitive cell. Aa a result of strong distortions of $NiO_6$, this phase has very low symmetry.

The corresponding Brillouin zone of the primitive cell of monoclinic $La_4Ni_3O_{10}$ with the $P2_1/c$ space group is plotted in Fig. 1(b) in which high symmetry points are denoted. To better understand the energy band, we should firstly know the little point group of various high symmetry points. The little point groups of points Γ (0, 0, 0), Z (0, 0.5, 0), B (0, 0, 0.5), and D (0, 0.5, 0.5) located in the $k_x = 0$ plane are all $C_{2h}$ (2/m). The point in the B-D direction is denoted as point V (0, u, 0.5) with symmetry of $C_2$ (2). The point group for Λ (0, u, 0) located at the Γ-Z direction is also $C_2$. Both Y (0.5, 0, 0) and $A_0$ (0.5, 0, 0) have symmetry $C_{2h}$. There is also an important point F (v, 0, u) in the $k_y = 0$ plane with the point group $C_s$ (m). The point located at the Γ-D direction is a general point (GP) with symmetry $C_1$ (1). These results are collected in in Table I.

The point group $C_{2h}$ has four group elements: identity $E$, a two-fold rotation $c_{2y}$

around the *y* axis, inversion *I* and a mirror $\sigma_h = Ic_{2y}$. Two group generators of the point group C$_{2h}$ are $c_{2y}$ and *I* with matrix forms:

$$R(c_{2y}) = \begin{pmatrix} -1 & 0 & 0 \\ 0 & 1 & 0 \\ 0 & 0 & -1 \end{pmatrix},$$

$$R(I) = \begin{pmatrix} -1 & 0 & 0 \\ 0 & -1 & 0 \\ 0 & 0 & -1 \end{pmatrix}. \quad (1)$$

Each element of the point group C$_{2h}$ forms a class. Hence C$_{2h}$ contains four classes. As a result, there are four one-dimensional irreducible representations $\Gamma_1^+$, $\Gamma_1^-$, $\Gamma_2^+$ and $\Gamma_2^-$. The character table of point group C$_{2h}$ is given in Table II.

From knowledge of group theory, we know that C$_2$, C$_s$ and C$_1$ are all subgroups of C$_{2h}$. Both C$_2$ and C$_s$ have two group elements. C$_2$ contains two group elements: identity *E* and $c_{2y}$. Two group elements for C$_s$ are *E* and a mirror $\sigma_h$. For both C$_2$ and C$_s$, each element forms a class. Thus, each of them has two one-dimensional irreducible representations $\Gamma_1$ and $\Gamma_2$. The point group C$_1$ only contains one element *E* and thus only has one one-dimensional irreducible representation $\Gamma_1$. According to these symmetry properties and Brillouin zone shown in Fig. 1(b), we may expect that electronic energy bands in the $k_y = 0$ and $k_x = 0$ planes might be very different. All these symmetry information will help us to understand electronic energy bands of monoclinic La$_4$Ni$_3$O$_{10}$.

**B. The energy gap in monoclinic La$_4$Ni$_3$O$_{10}$**

Now we analysis the origin of the tiny energy gap in monoclinic La$_4$Ni$_3$O$_{10}$ discovered in previous works. In Fig. 2 we plot electronic energy bands of monoclinic La$_4$Ni$_3$O$_{10}$ (primitive cell) with the $P2_1/c$ space group along high-symmetry directions of the Brillouin zone calculated with the above noted experimental lattice parameters. Our results are identical to previous experimental and theoretical works [6,7]. Energy bands are projected to Ni $3d_{x^2-y^2}$ (red) and Ni $3d_{z^2}$ (blue) orbitals. There is a tiny energy gap around the Brillouin zone center $\Gamma$ point near the fermi level E$_F$, which is formed by energy bands with the Ni $3d_{z^2}$ orbital character. This tiny energy gap has been confirmed experimentally [6] and is the important prominent feature of monoclinic La$_4$Ni$_3$O$_{10}$.

As noted above, the point group of the Brillouin zone center Γ point is C$_{2h}$ with four irreducible representations $\Gamma_{1,2}^{+,-}$. Based on symmetry analysis, we find that energy bands forming the gap at the Γ point transform according to the $\Gamma_2^+$ and $\Gamma_2^-$ irreducible representations of C$_{2h}$, as shown in Fig. 2. From above section we also know that, the momentum $k$ point located at the Γ-Z direction is denoted as Λ with symmetry C$_2$ having irreducible representations $\Gamma_1$ and $\Gamma_2$ (denoted as $\Lambda_1$ and $\Lambda_2$ now) and the point located at the Γ-D direction is GP with symmetry C$_1$ having an irreducible representation $\Gamma_1$ (denoted as GP$_1$ now). Compatibility relations in the group theory give the symmetry of energy bands at nearby $k$ points. According to compatibility relations between Γ and Z (GP), both $\Gamma_2^+$ and $\Gamma_2^-$ bands at the Γ point become $\Lambda_2$ bands at the Λ point or GP$_1$ bands at the GP point. These results are also displayed in Fig. 2. Since energy bands with the same symmetry character cannot cross, there are always energy gaps at Λ and GP points. This is usually called the anti-crossing in condensed matter physics. Thus, the tiny energy gap in monoclinic La$_4$Ni$_3$O$_{10}$ origins from energy bands having symmetry $\Gamma_2^+$ and $\Gamma_2^-$ with opposite parity at the Brillouin zone center Γ point. Only from these symmetry properties of energy bands near the Fermi level E$_F$ can we get a better understanding of the tiny energy gap.

We now construct a low energy effective Hamiltonian to describe the tiny energy gap around the Brillouin zone center Γ point in monoclinic La$_4$Ni$_3$O$_{10}$. We need two generators $c_{2y}$ and $I$ of the point group C$_{2h}$. From knowledge of characters of the point group C$_{2h}$ given in Table II, the symmetry representations in $\Gamma_2^+$ and $\Gamma_2^-$ bands are given by

$$D(c_{2y}) = \begin{pmatrix} -1 & 0 \\ 0 & -1 \end{pmatrix},$$
$$D(I) = \begin{pmatrix} 1 & 0 \\ 0 & -1 \end{pmatrix}. \tag{2}$$

The theory of invariant gives the following constraint on $k \cdot p$ model [15]

$$D(R)H(k)D^\dagger(R) = H(R(k)), \tag{3}$$

where $D(R)$ is the representative matrix of operator $R$. Based on these we obtain the $k \cdot p$ effective Hamiltonian (to second order in $k$) describing the tiny energy gap formed by $\Gamma_2^+$ and $\Gamma_2^-$ bands around the Fermi level E$_F$ in monoclinic La$_4$Ni$_3$O$_{10}$

$$H(k) = \varepsilon(k) + \begin{pmatrix} M(k) & (\alpha_3 - i\alpha_4)k_y \\ (\alpha_3 + i\alpha_4)k_y & -M(k) \end{pmatrix}, \quad (4)$$

with $\varepsilon(k) = \alpha_1 + \alpha_5 k_x^2 + \alpha_7 k_y^2 + \alpha_9 k_z^2 + \alpha_{11} k_x k_z$, $M(k) = \alpha_2 + \alpha_6 k_x^2 + \alpha_8 k_y^2 + \alpha_{10} k_z^2 + \alpha_{12} k_x k_z$ and $\alpha_i (= 1 \sim 12)$ are materials dependent parameters. We have eigenenergies:

$$E(k) = \varepsilon(k) \pm \sqrt{(\alpha_3^2 + \alpha_4^2)k_y^2 + M(k)^2} . \quad (5)$$

These energy bands show strong anisotropy in $k_{x,y,z}$ directions, which comes from low crystal symmetry of monoclinic La$_4$Ni$_3$O$_{10}$ shown in Fig. 1(a). The off-diagonal $k_y$ term in the Hamiltonian results from the group generator two-fold rotation $c_{2y}$ of point group C$_{2h}$ reflecting its different properties from $k_x$ and $k_z$ directions. The tiny energy gap is related to terms in the square root in (5). With this Hamiltonian we can have more accurate understanding of energy bands formed with Ni $3d_{z^2}$ orbitals. In the next section, we will investigate the anisotropic properties of energy bands in monoclinic La$_4$Ni$_3$O$_{10}$.

### C. Dirac points

From above two discussions we know that monoclinic La$_4$Ni$_3$O$_{10}$ has strong anisotropic symmetry properties. To having complete knowledge of energy bands in the entire Brillouin zone, in Fig. 3(a) we plot the two-dimensional energy bands at the $k_x = 0$ plane (Z-Γ-B-D) contributed by the Ni $3d_{z^2}$ orbital near the Fermi level E$_F$ around the Γ point. The energy gap can be found clearly. Amazingly, there are crossing points in k points located at the Γ-B ($k_z$) direction. Energy bands around the crossing point are shown in Fig. 3(b). In Fig. 4(a) we plot the electronic energy bands along high-symmetry B-Γ-A$_0$ directions of the Brillouin zone. As discussed above, $k$ points in the $k_y = 0$ plane are denoted as F (v, 0, u) with symmetry C$_s$ having two one-dimensional irreducible representations F$_1$ and F$_2$. Hence points between Γ and B (A$_0$) are all F. From compatibility relations, $\Gamma_2^+$ and $\Gamma_2^-$ bands at the Γ point become $F_2$ and $F_1$ bands, respectively, at the F point. According to knowledge of group theory, energy bands with different symmetry characters are allowed to cross. As a result, $F_2$ and $F_1$ bands cross at the F point as can be found in Fig. 4(a). Such crossing point is called the Dirac point like that in graphene [16,17]. Therefore, except the tiny energy gap, Ni $3d_{z^2}$ orbitals in monoclinic $La_4Ni_3O_{10}$ also form Dirac points. The energy bands in Γ-Z ($k_y$) and Γ-B ($k_z$) directions have very different properties, which comes from their different symmetry properties as Λ and F points having point groups $C_2$ and $C_s$ respectively. The Dirac point is protected by the mirror symmetry $\sigma_h$ in $C_s$. There is a linear energy-momentum relation around Dirac points as can be found in Fig. 3(b).

To see electronic energy bands in the $k_x$ direction (Γ-Y), in Fig. 4(b) we plot electronic energy bands along high-symmetry B-Γ-Y directions of the Brillouin zone shown with symmetry characters. The $k$ point in the Γ-Y direction is also F (v, 0, u), in which energy bands having $F_2$ and $F_1$ symmetry. Different from energy bands in the Γ-B direction, $F_2$ and $F_1$ bands do not crossing along the Γ-Y direction. In Fig. 5 we plot two-dimensional energy bands at the $k_y = 0$ plane (Y-Γ-B-$A_0$) contributed by the Ni $3d_{z^2}$ orbital near the Fermi level $E_F$ around the Γ point. The up band is very flat near Γ. While the down band looks like the surface of a saddle. Anisotropic band crossing characters between two bands can be found clearly. These results demonstrate that energy bands of monoclinic $La_4Ni_3O_{10}$ have strong anisotropy. Such character origins from anisotropy of its atomic structures. As can be found in Fig. 1(a), the distorted Ni-O planes tilt along the crystal lattice *a*. This leads that energy bands in $k_x$ and $k_z$ shown in Fig. 4(a) and 4(b) have very different dispersions. The strong distortions of $NiO_6$ results in very low crystal symmetry.

Previous experiments mainly focus on energy bands in the Z-Γ-D direction and thus do not find the Dirac points. In future ARPES experiments, Dirac points can be observed directly by focusing on energy bands in the B-Γ-$A_0$ direction. Both the tiny energy gap and Dirac points are important electronic characters of monoclinic $La_4Ni_3O_{10}$. These results also indicate the special role of the Ni $3d_{z^2}$ orbital.

Different physical properties of energy bands along $k_x$, $k_y$ and $k_z$ directions in monoclinic $La_4Ni_3O_{10}$ can also be understood from our Hamiltonian in (5). If $k_y = 0$, the off-diagonal term in Hamiltonian $H(k)$ disappears and thus $H(k)$ is a diagonal matrix. For $k_x = k_z = 0$, we have:

$$E(k_y) = \alpha_1 + \alpha_7 k_y^2 \pm \sqrt{(\alpha_3^2 + \alpha_4^2)k_y^2 + (\alpha_2 + \alpha_8 k_y^2)^2} \ . \tag{6}$$

While for $k_x = k_y = 0$, we have:

$$E(k_z) = \alpha_1 + \alpha_9 k_z^2 \pm (\alpha_2 + \alpha_{10} k_z^2) \ . \tag{7}$$

These show very different energy-momentum relations. Hence energy bands in $k_y$ (Γ-Z) and $k_z$ (Γ-B) with very different crossing properties are expected. This is consistent with the appearance of tiny energy gap and Dirac point in Fig. 2 and Fig. 4(a) respectively. Low crystal symmetry of monoclinic $La_4Ni_3O_{10}$ makes its energy bands having anisotropic properties. In a word, symmetry determines different dispersing electronic excitations in momentum space.

## IV. CONCLUSIONS

In summary, we have investigated electronic structures of monoclinic $La_4Ni_3O_{10}$ detailly. Based on group theory, we find that the tiny energy gap observed in previous experimental works origins from energy bands with opposite parity at the Brillouin zone center. In addition, previously unknown Dirac points in other momentum directions are also discovered. Energy bands of monoclinic $La_4Ni_3O_{10}$ have strong anisotropy due to its low crystal symmetry. Our works help to understand electronic characters of monoclinic $La_4Ni_3O_{10}$.

## ACKNOWLEDGMENTS

This work was supported by the Advanced Talents Incubation Program of the Hebei University (Grants No. 521000981423, No. 521000981394, No. 521000981395, and No. 521000981390), the Natural Science Foundation of Hebei Province of China (Grants No. A2021201001 and No. A2021201008), the National Natural Science Foundation of China (Grants No. 12104124 and No. 12274111), the Central Guidance on Local Science and Technology Development Fund Project of Hebei Province

TABLE I. The little point groups for high symmetry $k$ points in Brillouin zone of primitive cell of monoclinic $La_4Ni_3O_{10}$ with space group $P2_1/c$.

| $k$ point | Little group | $k$ point | Little group |
|---|---|---|---|
| Γ (0, 0, 0) | $C_{2h}$ (2/m) | Λ (0, u, 0) | $C_2$ (2) |
| Z (0, 0.5, 0) | $C_{2h}$ (2/m) | Y (0.5, 0, 0) | $C_{2h}$ (2/m) |
| B (0, 0, 0.5) | $C_{2h}$ (2/m) | $A_0$ (0.5, 0, 0) | $C_{2h}$ (2/m) |
| D (0, 0.5, 0.5) | $C_{2h}$ (2/m) | F (v, 0, u) | $C_s$ (m) |
| V (0, u, 0.5) | $C_2$ (2) | | |

TABLE II. Table of characters for point group $C_{2h}$ (2/m).

| | $E$ | $C_{2y}$ | $I$ | $\sigma_h$ |
|---|---|---|---|---|
| $\Gamma_1^+$ ($A_g$) | 1 | 1 | 1 | 1 |
| $\Gamma_1^-$ ($A_u$) | 1 | 1 | −1 | −1 |
| $\Gamma_2^+$ ($B_g$) | 1 | −1 | 1 | −1 |
| $\Gamma_2^-$ ($B_u$) | 1 | −1 | −1 | 1 |

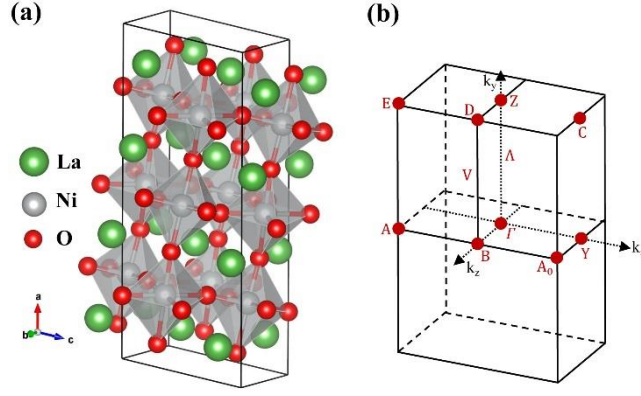

FIG. 1. (a) Atomic structures of monoclinic La$_4$Ni$_3$O$_{10}$ with the space group $P2_1/c$. (b) The Brillouin zone of the primitive cell of monoclinic La$_4$Ni$_3$O$_{10}$ shown with high symmetry points. The F point is located at (v, 0, u).

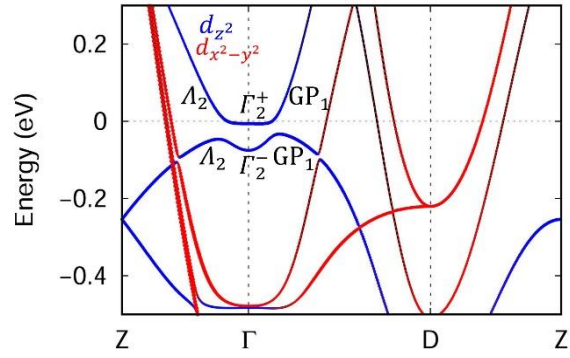

FIG. 2. First-principles calculated electronic band dispersions of monoclinic La$_4$Ni$_3$O$_{10}$ with the space group $P2_1/c$. The points between $\Gamma$ and Z (D) are denoted as $\Lambda$ (GP). The color lines indicate the projection to the Ni $3d_{x^2-y^2}$ (red) and Ni $3d_{z^2}$ (blue) states. Symmetry characters of energy bands are also shown.

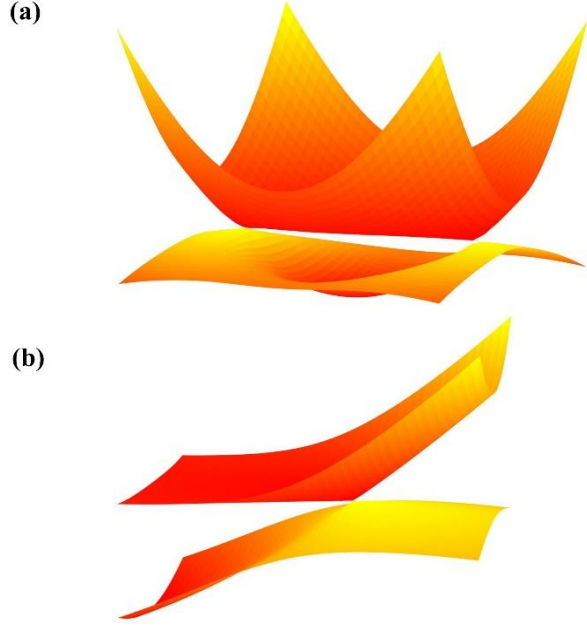

FIG. 3. (a) Two-dimensional energy bands of monoclinic $La_4Ni_3O_{10}$ at the $k_x = 0$ plane contributed by the Ni $3d_{z^2}$ orbital near the Fermi level $E_F$ around the Γ point. (b) Two-dimensional energy bands around the crossing point (Dirac point).

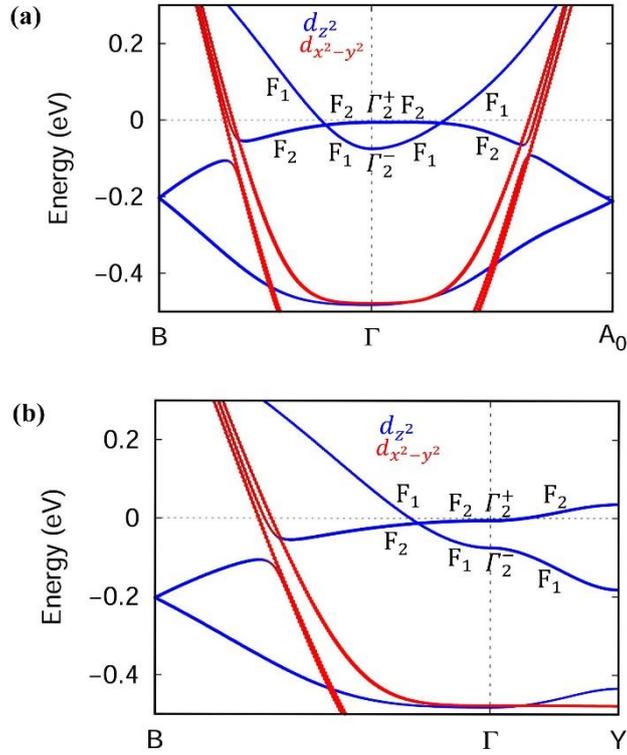

FIG. 4. First-principles calculated electronic band dispersions of monoclinic $La_4Ni_3O_{10}$ along high-symmetry (a) B-Γ-$A_0$ and (b) B-Γ-Y directions with the F (v, 0, u) point denoted. The color

lines indicate the projection to the Ni $3d_{x^2-y^2}$ (red) and Ni $3d_{z^2}$ (blue) states. Symmetry characters of energy bands are also shown.

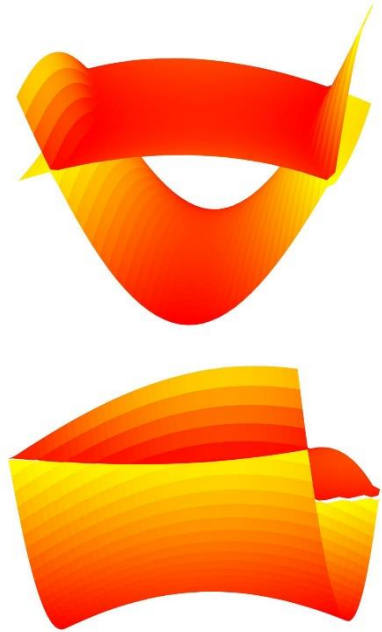

FIG. 5. Two-dimensional energy bands of monoclinic La$_4$Ni$_3$O$_{10}$ at the $k_y = 0$ plane contributed by the Ni $3d_{z^2}$ orbital near the Fermi level E$_F$ around the $\Gamma$ point.